\documentclass[twocolumn,showpacs,prb]{revtex4}

\usepackage{amsmath}
\usepackage{amssymb}
\usepackage{epsfig}
\usepackage{pstricks}

\def\Unitmatrix{\mbox{1\hspace{-.25em}I}}

\begin{document}

\title{Chiral symmetry restoration in (2+1)-dimensional $QED$ 
with a Maxwell-Chern-Simons term at finite temperature}
\author{Raoul Dillenschneider}
\email[E-mail address : ]{rdillen@lpt1.u-strasbg.fr}
\author{Jean Richert}
\email[E-mail address : ]{richert@lpt1.u-strasbg.fr}
\affiliation{
Laboratoire de Physique Th\'eorique, UMR 7085 CNRS/ULP,\\
Universit\'e Louis Pasteur, 67084 Strasbourg Cedex,\\ 
France} 

\date{\today}

\begin{abstract}
We study the role played by a Chern-Simons contribution to the action in the 
$QED_3$ formulation of a two-dimensional Heisenberg model of quantum spin 
systems with a strictly fixed site occupation at finite temperature. We show 
how this contribution affects the screening of the potential which acts between
spinons and contributes to the restoration of chiral symmetry in the spinon 
sector. The constant which characterizes the Chern-Simons term can be related to
the critical temperature $T_c$ above which the dynamical mass goes to zero.
\end{abstract}


\maketitle

\section{Introduction}

Quantum Electrodynamics $QED_{(2+1)}$ is a common framework aimed to 
describe strongly correlated systems such as quantum spin systems in $2$ 
space and $1$ time dimension, as well as related specific phenomena like
high-$T_c$ superconductivity 
\cite{Tesanovic-02,GhaemiSenthil-05, Morinari-05, LeeNagaosaWen-04}. 
Indeed, a gauge field formulation of the antiferromagnetic Heisenberg 
model in $d = 2$ dimensions leads to a $QED_3$ action for spinons,
see f. i. Ghaemi and Senthil  \cite{GhaemiSenthil-05}, Morinari 
\cite{Morinari-05} and also \cite{CondMat0602}.

We consider the $\pi$-flux state approach introduced by Affleck and 
Marston \cite{aff1,aff2}. The site-occupation of the system by a single
particle is enforced by means of a procedure which was suggested by Popov and 
Fedotov \cite{PRB,CondMat0602,Popov-88}.

Here we concentrate on the behaviour of the spinon mass at finite
temperature which is dynamically generated by an $U(1)$ gauge field when the
action contains a Chern-Simons (CS) term. Solving the Schwinger-Dyson equation 
for the spinon field we show below how the critical temperature $T_c$ (above 
which the dynamical mass vanishes) is affected by the CS term.

The outline of the paper is as follows. The first section sketches the main
steps of the $QED_3$ formulation of the $2d$ Heisenberg model. Section 
\ref{SectionMCS} introduces the CS term and justifies its presence. In section 
\ref{Section4} we show and comment the chiral symmetry restoration of the spinon
field in the presence of this term.

\section{From the Heisenberg interaction to the $\pi$-flux Dirac action}

Heisenberg quantum spin Hamiltonians of the type

\begin{equation}
H = -\frac {1}{2}\sum_{i,j} J_{ij} \vec S_{i} \vec S_{j}
\label{eq1}
\end{equation}

\noindent
with antiferromagnet coupling $\{J_{ij}\} < 0$ can be mapped onto Fock space 
by means of the transformation 
$S^{+}_{i} = f^{\dagger}_{i, \uparrow} f_{i, \downarrow}$,
$S^{-}_{i} = f^{\dagger}_{i, \downarrow} f_{i, \uparrow}$ and
$S^{z}_{i} = \frac{1}{2} (f^{\dagger}_{i,\uparrow} f_{i,\uparrow} - 
f^{\dagger}_{i,\downarrow} f_{i,\downarrow})$ 
where $\{f^\dagger_{i, \sigma}, f_{i, \sigma}\}$ are anticommuting 
fermion operators which create and annihilate spinons with $\sigma=\pm 1/2$.
The projection onto Fock space is exact when the number of fermions per 
lattice site verifies 
$\underset{\sigma=\pm 1/2}{\sum} f^\dagger_{i,\sigma} f_{i,\sigma}=1$.
This can be enforced by using the Popov and Fedotov procedure
\cite{PRB,CondMat0602,Popov-88} which introduces the imaginary chemical
potential $\mu = i \pi / 2 \beta$ at temperature $\beta^{-1}$ adding the 
term $\mu N$ to the expression given by \eqref{eq1}

\begin{equation*}
H = -\frac {1}{2}\sum_{i,j} J_{ij} \vec S_{i} \vec S_{j} - \mu N
\end{equation*}

\noindent
where $N = \underset{i,\sigma}{\sum} f^\dagger_{i,\sigma} f_{i,\sigma}$ 
counts the number of fermions in the spin system.

In 2$d$ space the Heisenberg Hamiltonian given by Eq.\eqref{eq1} can 
be written in terms of composite non-local operators $\{{\cal D}_{ij}\}$ 
("diffusons") \cite{aue,PRB} defined as 

\begin{eqnarray*}
{\mathcal D}_{ij} = f_{i, \uparrow}^{\dagger} f_{j, \uparrow} + 
f_{i, \downarrow}^{\dagger} f_{j, \downarrow}
\end{eqnarray*}

\noindent
If the coupling strengths are fixed as 

\begin{eqnarray*}
J_{ij} = J 
\underset{\vec{\eta}}{\sum} \delta \left(\vec{r}_i - \vec{r}_j \pm \vec{\eta} 
\right)
\end{eqnarray*}

\noindent
where $\vec{\eta}$ is a lattice vector $\{a_1,a_2\}$ in the $\vec {Ox}$  
and $\vec {Oy} $ directions the Hamiltonian takes the form

\begin{eqnarray}
H = - J \sum_{<ij>} (\frac{1}{2} {\cal D}^{\dagger}_{ij} {\cal D}_{ij} - 
\frac{n_i}{2} + \frac{n_i  n_j}{4})  - \mu N
\label{eq2} 
\end{eqnarray}

\noindent
where $i$ and $j$  are nearest neighbour sites.

The number operator products $\{n_i  n_j\}$ in Eq.\eqref{eq2} are quartic 
in terms of creation and annihilation operators in Fock space. 
In principle the formal treatment of these terms requires the introduction
of a Hubbard-Stratonovich (HS) transformation. 
One can however show that the presence of this term has no
influence on the results obtained from the partition function. 
Indeed both $\{n_i\}$ and $\{n_i  n_j\}$ lead to constant contributions 
under the exact site-occupation constraint and hence are of no importance for 
the physics described by the  Hamiltonian \eqref{eq2}.
As a consequence we leave them out from the beginning.  

Using a HS transformation in order to reduce the first term in
Eq. \eqref{eq2} from quartic to quadratic order in the fermion operators 
$f^\dagger$ and $f$ the Heisenberg Hamiltonian reads

\begin{eqnarray}
\mathcal{H} = \frac{2}{|J|}\underset{<ij>}{\sum} \bar{\Delta}_{ij} \Delta_{ij} 
+\underset{<ij>}{\sum} \left[ \bar{\Delta}_{ij} {\cal D}_{ij} +
\Delta_{ij} {\cal D}^{\dagger}_{ij} \right]  - \mu N
\notag \\
\label{eq3}
\end{eqnarray}

\noindent
where $\left\{ \Delta_{ij} \right\}$ are the HS auxiliary fields. 
At this point no approximation has been made and Eq.\eqref{eq3} is exact.

\begin{figure}
\center
\epsfig{file=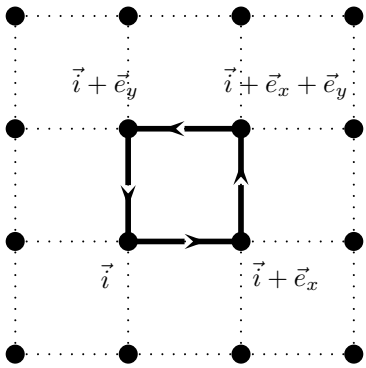}
\caption{Plaquette ($\Box$) on a two dimensional spin lattice.
$\vec {e}_{x}$ and $\vec {e}_{y}$ are the unit vectors along the directions 
$\vec {Ox}$ and $\vec {Oy}$ starting from site $\vec i$ on the lattice}
\label{Fig1}
\end{figure}

The fields $\Delta_{ij}$ can be chosen as complex quantities 
$\Delta_{ij}=|\Delta| e^{i\phi_{ij}}$. This parametrization introduces gauge
fields $\phi_{ij}$ which are defined on the 
square plaquette shown in figure \ref{Fig1} and can be decomposed
into a mean field part and a fluctuation contribution 
$\phi_{ij} = \phi_{ij}^{mf} + \delta \phi_{ij}$. The amplitude $|\Delta|$ too
may contain a mean-field and a fluctuating contribution.
In the following we assume that at low energy the essential quantum 
fluctuation contributions are generated by the gauge field $\delta \phi_{ij}$ 
and neglect the amplitude fluctuations in the sequel.

The $\phi_{ij}^{mf}$'s are fixed on the plaquette in such a way that

\begin{eqnarray*}
\phi^{mf} = \sum_{(ij) \in \Box}\phi_{ij}^{mf}
\end{eqnarray*}

\noindent  
where $\phi^{mf}$ is taken to be constant.

In order to implement the $SU(2)$ invariance in \eqref{eq1} at the level of 
the mean-field Hamiltonian \eqref{eq3} we follow \cite{wen,aff1,aff2,awe,lee} 
and introduce the configuration

\begin{eqnarray*}
\phi_{ij}^{mf}=
\begin{cases}
\frac{\pi}{4}(-1)^i, \text{if } \vec{r}_j=\vec{r}_i+\vec{e}_x \\
-\frac{\pi}{4}(-1)^i, \text{if } \vec{r}_j=\vec{r}_i+\vec{e}_y \\
\end{cases}
\end{eqnarray*}

\noindent
Then the total flux through the fundamental plaquette is such that 
$\phi^{mf} = \pi$ which guarantees the $SU(2)$ symmetry \cite{Marston}.

Under these conditions the Hamiltonian \eqref{eq3} goes over to the $\pi$-flux 
mean-field Hamiltonian

{\small
\begin{eqnarray}
\mathcal{H}_{MF}^{(PFP)} = \mathcal{N} z \frac{\Delta^2}{|J|}
+ \underset{\vec{k} \in SBZ}{\sum} \underset{\sigma}{\sum}
&&
(
f^\dagger_{\vec{k},\sigma} \,
f^\dagger_{\vec{k}+\vec{\pi},\sigma}
)
[ \widetilde{H} ]
\left(
\begin{array}{c}
f_{\vec{k},\sigma} \\
f_{\vec{k}+\vec{\pi},\sigma}
\end{array}
\right)
\notag \\
\label{Chapter5eq1}
\end{eqnarray}}

\noindent
with

\begin{eqnarray*}
\left[ \widetilde{H} \right] =
\left[
\begin{array}{cc}
-\mu + \Delta \cos \left( \frac{\pi}{4}\right) \gamma_{k_x,k_y} &
-i \Delta \sin \left( \frac{\pi}{4} \right) \gamma_{k_x,k_y+\pi} \\
+i \Delta \sin \left( \frac{\pi}{4} \right) \gamma_{k_x,k_y+\pi} &
-\mu - \Delta \cos \left( \frac{\pi}{4} \right) \gamma_{k_x,k_y} 
\end{array}
\right]
\end{eqnarray*}

\noindent
where the $\gamma_{\vec{k}}$'s are defined by 
$\gamma_{\vec{k}} = \underset{\vec{\eta}}{\sum} 
e^{i \vec{k}.\vec{\eta}} = 2 \left( \cos k_x a_1  + \cos k_y a_2 \right)$.
The eigenvalues of $\mathcal{H}_{MF}^{(PFP)}$ read 
$\omega^{(PFP)}_{(\pm),\vec{k},\sigma} = - \mu \pm 2\Delta
\sqrt{\cos^2(k_x) + \cos^2(k_y)}$.

\begin{figure}
\center
\epsfig{file=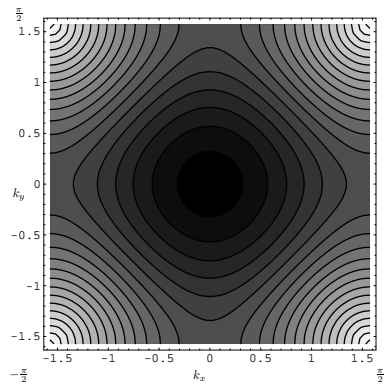,width=5cm}
\caption{The contour representation of the energy spectrum 
$\omega^{(PFP)}_{(-)\vec{k},\sigma} + \mu = - 2\Delta
\sqrt{\cos^2(k_x) + \cos^2(k_y)}$ for $k_x$ and $k_y$ belonging to 
$\left[-\frac{\pi}{2},\frac{\pi}{2}\right]$ and 
showing the presence of the nodal points $(\pm \frac{\pi}{2},\pm 
\frac{\pi}{2})$ where the energy is equal to zero.}
\label{Chapter5figEnergy}
\end{figure}

We are interested in the low energy behaviour of the quantum spin system 
described by \eqref{Chapter5eq1} in the neighbourhood of the nodal
points $\left( k_x = \pm \frac{\pi}{2}, k_y = \pm \frac{\pi}{2} \right)$ 
where the energy gap $\left( \omega^{(PFP)}_{(+),\vec{k},\sigma} -
\omega^{(PFP)}_{(-),\vec{k},\sigma} \right)$ vanishes. 
Fig.\ref{Chapter5figEnergy} shows the contour plot of the energy spectrum 
$\omega^{(PFP)}_{(-),\vec{k},\sigma}$ and locates the nodal points. 
We linearize the energies in the neighbourhood of these points.

Following \cite{GhaemiSenthil-05,Morinari-05,CondMat0602} the spin liquid
Hamiltonian \eqref{Chapter5eq1} at low energy can be described in terms of 
four-component Dirac spinons in the continuum limit. The Dirac action of 
this spin liquid in (2+1) dimensions including the phase fluctuations
$\delta \phi_{ij}$ around the $\pi$-flux mean field phase $\phi_{ij}^{mf}$ 
has been derived in \cite{CondMat0602} and reads

\begin{eqnarray}
S_{E} = \int_0^\beta \int d^2\vec{r}
&\Bigg\{&
- \frac{1}{2} a_\mu \left[ \left(\Box \delta^{\mu \nu}
+ (1 - \lambda) \partial^\mu \partial^\nu \right) \right] a_\nu
\notag \\
&+& \underset{\sigma}{\sum}
\bar{\psi}_{\vec{r} \sigma} \left[
\gamma_\mu \left( \partial_\mu - i g a_\mu \right)
\right] \psi_{\vec{r} \sigma}
\Bigg\}
\label{SpinonGaugeAction}
\end{eqnarray}

\noindent
$\psi$ is the 4-dimensional Dirac spinon field  

\begin{eqnarray*}
\psi_{\vec{k} \sigma} = \left(
\begin{array}{c}
f_{1 a, \vec{k} \sigma} \\
f_{1 b, \vec{k} \sigma} \\
f_{2 a \vec{k} \sigma} \\
f_{2 b \vec{k} \sigma}
\end{array}
\right)
\end{eqnarray*}

\noindent
where $f^\dagger_{1,\vec{k},\sigma}$ and $f_{1,\vec{k},\sigma}$ 
($f^\dagger_{2,\vec{k},\sigma}$ and $f_{2,\vec{k},\sigma}$) 
are fermion creation and annihilation operators near the nodal points
$(\frac{\pi}{2},\frac{\pi}{2})$ ($(-\frac{\pi}{2},\frac{\pi}{2})$) 
of the momentum $\vec{k}$ and indices $a$ and $b$ characterize 
the rotated operators

\begin{eqnarray*}
\begin{cases}
f_{a,\vec{k},\sigma} = \frac{1}{\sqrt{2}}
\left(f_{\vec{k},\sigma} + f_{\vec{k}+\vec{\pi},\sigma}\right) \\
f_{b,\vec{k},\sigma} = \frac{1}{\sqrt{2}}
\left(f_{\vec{k},\sigma} - f_{\vec{k}+\vec{\pi},\sigma}\right)  
\end{cases}
\end{eqnarray*}

\noindent
The first term in \eqref{SpinonGaugeAction} originates from the $U(1)$ symmetry
transformation $\psi \rightarrow  e^{i g \theta} \psi$ which generates a gauge 
field $a_\mu = \partial_\mu \theta$. The constant $g$ in 
\eqref{SpinonGaugeAction} is the coupling strength between the gauge field 
$a_\mu$ and the Dirac spinons $\psi$. The first term corresponds to the 
``Maxwell'' term $-\frac{1}{4} f_{\mu \nu} f^{\mu \nu}$ of the gauge field 
$a_\mu$ where $f^{\mu \nu} = \partial^\mu a_\nu - \partial^\nu a_\mu$,
$\lambda$ is the parameter of the Faddeev-Popov gauge fixing term
$-\lambda \left(\partial^\mu a_\mu \right)^2$ \cite{Itzykson},   
$\delta^{\mu \nu}$ the Kronecker $\delta$, 
$\Box = \partial_\tau^2 + \vec{\nabla}^2$ the Laplacian in Euclidean 
space-time. This form of the action originates from a shift of the imaginary 
time derivation $\partial_\tau \rightarrow \partial_\tau + \mu$ where $\mu$
is the chemical potential introduced above. It leads to a new definition of the
Matsubara frequencies of the fermion fields \cite{Popov-88} $\psi$ which then 
read $\widetilde{\omega}_{F,n} = \omega_{F,n} - \mu/i = \frac{2 \pi}{\beta} 
(n + 1/4)$.

\section{Maxwell-Chern-Simons action at finite temperature
\label{SectionMCS}}

\subsection{Justification and implementation}

As shown by Marston \cite{Marston}, only gauge configurations of the flux 
states belonging to $Z_2$ symmetry ($\pm \pi$) are allowed. Hence the
flux through the plaquette is restricted to 
$\phi = \phi^{mf} + \delta \phi =\left\{ 0,\pm \pi \right\}$.
In order to remove ``forbidden'' $U(1)$ gauge configurations of the
antiferromagnet Heisenberg model ($\phi \neq \pm \pi$) a CS term should be 
included in the $QED_3$ action in order to fix the total flux through a 
plaquette. This leads to the  Maxwell-Chern-Simons (MCS) action in Euclidean 
space

\begin{eqnarray}
S_{E} &=& \int_0^\beta \int d^2\vec{r}
\Bigg\{
- \frac{1}{2} a_\mu \Big[ \left(\Box \delta^{\mu \nu}
+ (1 - \lambda) \partial^\mu \partial^\nu \right) 
\notag \\
&&
+ i \kappa \varepsilon^{\mu \rho \nu} \partial_\rho  \Big] a_\nu
+ \underset{\sigma}{\sum}
\bar{\psi}_{\vec{r} \sigma} \left[
\gamma_\mu \left( \partial_\mu - i g a_\mu \right)
\right] \psi_{\vec{r} \sigma}
\Bigg\}
\notag \\
\label{MCSAction}
\end{eqnarray}

\noindent
The implementation of the CS action

\begin{equation}
S^{CS}_{E} = \int_0^\beta d\tau \int d^2\vec{r} \left( i \frac{\kappa }{2} 
\varepsilon^{\mu \rho \nu} a_\mu \partial_\rho   a_\nu \right)
\end{equation}

\noindent
introduces a new constant $\kappa$. We show below that this constant can be 
fixed to a definite value.  

From the above action \eqref{MCSAction}, the equation of motion of the gauge 
field in Minkowskian space

\begin{eqnarray}
\partial_\nu f^{\nu \mu} - 
\left(\kappa/2 \right) \varepsilon^{\mu \nu \rho} f_{\nu \rho} = 
- g \underset{\sigma}{\sum}\bar{\psi}_{\sigma} \gamma^{\mu} \psi_{\sigma}
\label{eqMotion}
\end{eqnarray}

\noindent
leads to a relation between a magnetic field and the 
CS coefficient \cite{Dunne,ItohKato}. 
If $\mathcal{B} = \partial_1 a_2 - \partial_2 a_1$ 
%(which derives from $a_\mu$'s is not to be confused with the real magnetic
%field $B = \partial_1 A_2 - \partial_2 A_1$ deriving from the real
%potential vector $A_\mu$) 
is chosen to be constant in such a way that the whole system experiences an 
homogeneous magnetic field the equation of motion \eqref{eqMotion} of the 
gauge field becomes

\begin{equation}
\kappa \mathcal{B} 
= -g \underset{\sigma}{\sum} <\psi^\dagger_\sigma \psi_\sigma >
\label{eqKappaB}
\end{equation}

\noindent
The gauge field $a_\mu$ is related to the phase $\theta \left( \vec{r} \right)$
of the spinon at site $\vec{r}$ through the gauge transformation 
$f_{\vec{r},\sigma} \rightarrow e^{i g \theta \left( \vec{r} \right)} 
f_{\vec{r},\sigma}$ which keeps the Heisenberg Hamiltonian \eqref{eq3} 
invariant. From the definition of $\psi$ one gets $\psi_{\vec{r} \sigma} 
\rightarrow e^{i g \theta \left( \vec{r} \right)}\psi_{\vec{r} \sigma}$.
It is clear that $\theta \left( \vec{r} \right)$ is the phase at the lattice 
site $\vec{r}$ and that $a_\mu\left( \vec{r}\right) = \partial_\mu \theta 
\left( \vec{r} \right)$. Hence the magnetic field $\mathcal{B}$ is then 
directly related to the flux $\phi$ through the plaquette shown in 
figure \ref{Fig1}

\begin{eqnarray}
\phi &=& g \underset{<i,j> \in \Box}{\sum} 
\left(
\theta \left( \vec{r_i} \right) - \theta \left( \vec{r_j} \right) 
\right) = g \int_{\Box} d\vec{l}.\vec{a}
\notag \\
&=&  g \int_{\Box} d\vec{\Omega}_{\Box}.\vec{\mathcal{B}} 
= g \Omega_{\Box} \mathcal{B} = \pi k
\label{eqFlux}
\end{eqnarray}

\noindent
where $\Omega_{\Box}$ is the surface of the plaquette and $k$ an integer. 
Here the flux is fixed to be to be equal to $\left\{0,\pm \pi \right\}$ 
\cite{Marston}.

Hence $\kappa$ can be fixed by the flux through the plaquette using
equations \eqref{eqFlux} and \eqref{eqKappaB}. 
Defining $\rho = \underset{\sigma}{\sum} <\psi_{\sigma}^\dagger 
\psi_{\sigma}>$ as the density of spinon one can indeed rewrite equation 
\eqref{eqKappaB}

\begin{eqnarray*}
\kappa = -\frac{g \rho}{\mathcal{B}} = \frac{g^2}{\pi} \frac{\mathcal{N}}{k}
\end{eqnarray*}

\noindent
where $\mathcal{N}$ is the number of spinons on the plaquette and $k$ is 
an integer, see \eqref{eqFlux}. Recalling that

\begin{eqnarray}
\nu = \frac{\text{number of particle}}{\text{number of flux quanta}} = 
\frac{\rho}{g |\mathcal{B}|/(2\pi)} = \frac{2 \mathcal{N}}{k} 
\end{eqnarray}

\noindent
is the filling factor of the Landau level \cite{NagaosaBook} in the Quantum
Hall Effect ($QHE$) one finally gets 

\begin{eqnarray}
\kappa = \frac{g^2}{2 \pi} \nu 
\end{eqnarray}

\noindent
The present analysis may suggest that the application of a real magnetic field 
to the spin system could allow to detect the presence of spinons through the 
Quantum Hall Effect \cite{NagaosaBook}. We leave this point for further 
investigations.

\subsection{The photon propagator at finite temperature}

In this subsection we construct the dressed photon propagator of $QED_3$ with 
an MCS term at finite temperature in order to gain information about the 
interaction between spinons $\psi$ and the implication of the CS term on the 
dynamical mass generation.

Integrating \eqref{MCSAction} 
over the fermion fields $\psi$, the partition function of the spin system 
$\mathcal{Z}\left[ \psi,a \right] = \int \mathcal{D}\left(\psi,a\right) 
e^{-S_{E}}$ with the action $S_{E}$ given by \eqref{MCSAction} leads 
to the pure gauge partition function

\begin{eqnarray}
\mathcal{Z}\left[ a \right] = 
\int \mathcal{D}\left( a \right) e^{ - S_{eff}\left[ a \right] }
\end{eqnarray}

\noindent
where the effective pure gauge field action $S_{eff}\left[ a \right]$ comes
in the form

\begin{eqnarray}
S_{eff} \left[ a \right] &=&
\int_0^\beta d\tau \int d^2\vec{r} 
\Bigg\{
- \frac{1}{2} a_\mu \Big[ 
\left(\Box \delta^{\mu \nu}
+ (1 - \lambda) \partial^\mu \partial^\nu \right) 
\notag \\
&&
+ i \kappa \varepsilon^{\mu \rho \nu} \partial_\rho
\Big] a_\nu
\Bigg\}
- \ln \, \det 
\left[  \gamma_\mu \left( \partial_\mu - i g a_\mu \right) \right]
\notag \\
\label{Chapter5eq10}
\end{eqnarray}

\noindent
One can develop the last term in the effective gauge field action 
$S_{eff}\left[ a \right]$ into a series and write

\begin{gather}
\ln \, \det \left[  \gamma_\mu \left( \partial_\mu - i g a_\mu \right) \right]
=
\notag \\
\ln \, \det G_{F}^{-1}
 - \overset{\infty}{\underset{n = 1}{\sum}}
\frac{1}{n} Tr \left[ i G_{F} \gamma^\mu a_\mu \right]^n
\label{Chapter5eq11}
\end{gather}

\noindent
where $G_{F}^{-1}(k-k^{'}) = i \frac{\gamma^\mu k_\mu}{(2\pi)^2\beta} 
\delta(k-k^{'})$ is the fermion Green function in the Fourier space-time with
$k = \left( \widetilde{\omega}_{F,n}, \vec{k} \right)$, hence 
$G_{F} =-i \frac{\gamma^\mu k_\mu}{k^2} (2\pi)^2\beta 
\delta\left( k - k^{'} \right)$. The first term on the r.h.s. of 
equation \eqref{Chapter5eq11} is independent of the gauge field 
$\{ a_\mu \}$. It can be removed from the series since we focus our attention
on pure gauge field terms. 
The first term proportional to the gauge field $n=1$ in the sum vanishes 
since $tr \, \gamma_\mu = 0$. Keeping only second order terms 
in order to stay with gaussian contributions to the fluctuations one gets the 
pure gauge action 

\begin{gather}
S_{eff}^{(2)}\left[ a \right] =
\int_0^\beta d\tau \int d^2\vec{r} 
\Bigg\{
- \frac{1}{2} a_\mu \Big[ 
\left(\Box \delta^{\mu \nu}
+ (1 - \lambda) \partial^\mu \partial^\nu \right) 
\notag \\
+ i \kappa \varepsilon^{\mu \rho \nu} \partial_\rho
\Big] a_\nu
\Bigg\}
+ \frac{g^2}{2 \beta} \underset{\sigma}{\sum} 
\underset{\omega_{F,1}}{\sum} \int \frac{d^2\vec{k}_1}{(2\pi)^2}.
\frac{1}{\beta} \underset{\omega_{F}^{''}}{\sum} \int \frac{d^2\vec{k^{''}}}
{(2\pi)^2}
\notag \\
\times
tr \Bigg[
\frac{\gamma^\rho k_{1,\rho}}{k_1^2} \gamma^\mu a_\mu(k_1-k^{''})
\frac{\gamma^{\eta} k^{''}_\eta}{{k^{''}}^2}
\gamma^\nu a_\nu \left(-(k_1 - k^{''}) \right)
\Bigg]
\notag \\
\label{Chapter5eq13}
\end{gather}

\noindent
The second term in equation \eqref{Chapter5eq13} has been worked out in 
\cite{CondMat0602}.  The whole action can be put into the form
 
\begin{eqnarray}
S_{eff}^{(2)}\left[ a \right]&=&
- \frac{g^{2}}{2 \beta}  
\underset{\omega_{B}}{\sum} \int \frac{d^2\vec{q}}{(2\pi)^2}
\notag \\
&& \times
a_\mu(-q) \left[ {\Delta^{(0)}_{E \mu \nu}}^{-1} 
+ \Pi_{\mu \nu}(q) \right] a_\nu(q)
\end{eqnarray}

\noindent
where ${\Delta^{(0)}_{E \mu \nu}} = \frac{1}{q^2 \left(q^2 + \kappa^2 \right)}
\left[q^2 \delta_{\mu \nu} - q_\mu q_\nu - \kappa \varepsilon_{\mu \nu \rho} 
q^\rho \right] + \frac{1}{\lambda} \frac{q_\mu q_\nu}{\left( q^2\right)^2}$ 
is the bare photon propagator in Euclidean space-time.
The one-loop vacuum polarization term \cite{CondMat0602} reads

\begin{eqnarray*}
\Pi_{\mu \nu} &=& \Pi_A A_{\mu \nu} + \Pi_B B_{\mu \nu}
 \\
&=& \left( \widetilde{\Pi}_1(q_\mu) + 
\widetilde{\Pi}_2(q_\mu) \right) A_{\mu \nu} + \widetilde{\Pi}_3(q_\mu) 
B_{\mu \nu}
\end{eqnarray*}

\noindent
where $A_{\mu \nu}$ and $B_{\mu \nu}$ are Lorentz invariant tensors given in
the Appendix and

\begin{eqnarray*}
\widetilde{\Pi}_1(q_\mu) &=& \frac{\alpha q}{\pi} \int_0^1 dx \sqrt{x(1-x)}
\frac{\sinh \beta q \sqrt{x(1-x)} }{D(X,Y)}
 \\
\widetilde{\Pi}_2(q_\mu) &=& \frac{\alpha m}{\beta}
\int_0^1 dx (1-2x) \frac{\cos 2 \pi x m}{D(X,Y)}
 \\
\widetilde{\Pi}_3(q_\mu) &=& \frac{\alpha}{\pi \beta} \int_0^1 dx \log 2 D(X,Y)
\end{eqnarray*}

\noindent
with $D(X,Y)= \cosh \left( \beta q \sqrt{x(1-x)} \right) + \sin (2\pi x m) $.
Here the photon momentum $q_\mu = \left(\omega_{B,m} = \frac{2 \pi m}{\beta}
,\vec{q} \right)$ with $\mu = \left\{ 0, 1, 2 \right\}$, $m$ is an integer and 
$\alpha = 2 g^2$ the coupling constant.

The finite-temperature dressed photon propagator in Euclidean 
space verifies the Dyson equation

\begin{eqnarray}
\Delta_{E \mu \nu}^{-1} &=& {\Delta_{E \mu \nu}^{(0)}}^{-1} + \Pi_{\mu \nu}
\label{Dyson}
\end{eqnarray}

\noindent
The inversion of equation \eqref{Dyson} leads to the dressed photon propagator
with the CS term at finite temperature

\begin{eqnarray}
\Delta_{E \mu \nu} &=&
\left[\left(q^2 + \Pi_A \right) A_{\mu \nu}
+ \left(q^2 + \Pi_B \right) B_{\mu \nu} - \kappa \varepsilon_{\mu \nu \rho}
q^\rho \right]
\notag \\
&&
/\left[\left(q^2 + \Pi_A \right)
\left(q^2 + \Pi_B \right) + \left(\kappa q \right)^2 \right] 
+ \frac{q_\mu q_\nu}{\lambda \left( q^2\right)^2}
\notag \\
\label{MCSpropagator}
\end{eqnarray}

\section{``Chiral'' symmetry restoration \label{Section4}}

The coupling of the gauge field  $a_{\mu}$ to the spinon field generates 
a mass for this field\cite{CondMat0602}. Chiral symmetry in four dimensions 
requires fermions to be massless. In this space a mass term 
$m \bar{\psi}{\psi}$ changes 
sign under chiral transformations generated by means of the Dirac matrix 
$\gamma_5$. Hence fermions must be massless in order to keep the action 
invariant. In three dimensions no real $\gamma_5$ matrix can be defined.
However embedding the (2+1)-dimensional space into a four dimensional space
two types of ``chiral'' symmetries can be defined
from $\gamma_3$ and $\gamma_5$ \cite{DoreyMavromatos,Appelquist1,Rausch}
where $\gamma_3$ and $\gamma_5$ are $4 \times 4$ matrices   
 
\begin{eqnarray*}
\gamma_3 =
\left(
\begin{array}{cc}
0 & \Unitmatrix \\
\Unitmatrix & 0
\end{array}
\right)
, \,
\gamma_5 = i
\left(
\begin{array}{cc}
0 &  \Unitmatrix \\
- \Unitmatrix & 0
\end{array}
\right)
\end{eqnarray*}

\noindent
which induce "chiral" transformations $e^{i g \theta \gamma_3}$ and 
$e^{i g \theta \gamma_5}$. In (2+1) dimensions the algebra is completed 
by

\begin{gather*}
\gamma^0 =
\left(
\begin{array}{cc}
\tau_3 & 0 \\
0 & -\tau_3
\end{array}
\right)
 ,\,
\gamma^1 =
\left(
\begin{array}{cc}
\tau_1 & 0 \\
0 & -\tau_1
\end{array}
\right)
 ,\,
\gamma^2 =
\left(
\begin{array}{cc}
\tau_2 & 0 \\
0 & -\tau_2
\end{array}
\right)
\end{gather*}

\noindent
where $\{\tau_i, i = 1,2,3\}$ are the Pauli matrices and Dirac matrices verify 
$\gamma_\mu \gamma_\nu + \gamma_\nu \gamma_\mu = 2 \delta_{\mu \nu}$ in 
Euclidean space.  

Appelquist et al. \cite{Appelquist1,Appelquist2} showed that at zero 
temperature the originally massless fermion can acquire a dynamical mass 
when the number 
$N$ ($ = \underset{\sigma}{\sum} \, 1$) of fermion flavors is lower than the 
critical value $N_c = 32/\pi^2$. Later Maris \cite{Maris} confirmed this 
result with $N_c \simeq 3.3$. Since we consider only spin-$1/2$ systems, $N=2$ 
and hence $N<N_c$.

At zero temperature the dynamical mass term is renormalized by the CS 
term $\frac{m(\kappa \neq 0)}{m(\kappa = 0)} = e^{\left[ - \frac{4 N}{N_c}.
\frac{\kappa^2}{(\alpha/16)^2} \right]}$ and even the critical value $N_c$
is affected as $\widetilde{N}_c = N_c \left[ 1 + (16 \kappa/\alpha)^2 \right]$ 
as shown by Hong and Park \cite{HongPark}.

Here we concentrate on the impact of the CS term on the dynamical
mass generation and show that chiral symmetry can be restored at finite
temperature. An explanation of the mechanism behind this symmetry restoration 
will also be given.

\subsection{Effective potential at finite temperature}

In the present theory mass is generated in two different ways. First, as shown
earlier, the massless photon induces a mass for the spinon through the coupling
of the two fields \cite{CondMat0602}. Second, the CS coefficient gives a mass 
to the "photon" (gauge field $a_{\mu}$), $m_{MCS} = \kappa$. This can be seen 
from the pure gauge equation of motion for the dual field 
$\widetilde{f}_\mu \equiv \frac{1}{2} \varepsilon^{\mu \nu \rho} f_{\nu \rho}$
\cite{Dunne}  

\begin{eqnarray*}
\left(\partial^\mu \partial_\mu + \kappa^2 \right) \widetilde{f}_\nu = 0
\end{eqnarray*}

\noindent
The massive photon induces the same effect (dynamical mass generation) at 
zero temperature \cite{HongPark}.

We show now how the photon mass $\kappa$ (the CS coefficient) affects the 
effective potential at finite temperature between two spinons.

The static effective potential between spinons with opposite charge $g$ is 
given by

\begin{eqnarray*}
V(R) &=& -g^2 \int_0^\beta d\tau \Delta_{00} \left(\tau,R \right) \\
&=& -\frac{g^2}{2\pi} \int \frac{d^2\vec{q}}{(2\pi)^2}
\Delta_{00} \left(q^0 = 0,\vec{q}e^{i \vec{q}.\vec{R}} \right) \\
&=& -\frac{g^2}{2\pi} \int_0^\infty dq.q. J_0(qR).\Delta_{00}\left(0,\vec{q} 
\right)
\end{eqnarray*}

\noindent
$J_0(qR)$ is the Bessel function of the first kind and

\begin{eqnarray*}
\Delta_{00} = \frac{1}{
\left(q^2 + \Pi_B(m=0) \right) + \frac{\left(\kappa q \right)^2}{
\left(q^2 + \Pi_A(m=0) \right)}}
\end{eqnarray*}

\noindent
At large distances $q \rightarrow 0$ the one-loop vacuum 
polarization parts become $\Pi_A(m=0) \underset{q \rightarrow 0}{=} 
q^2 .\frac{\alpha \beta} {12 \pi}$ and 
$\Pi_B(m=0) \underset{q \rightarrow 0}{=} \frac{\alpha}{\pi \beta} \ln 2$
where the integer $m$ is related to the photon energy, see above \eqref{Dyson}.
Hence the longitudinal part of the photon propagator $\Delta_{00}$ leads to 
the definition of a correlation length $\xi_\kappa$

\begin{eqnarray*}
\Delta_{00}\left(0,\vec{q} \right) = \frac{1}{
q^2 + \xi_\kappa^{-2}}
\end{eqnarray*}

\noindent
where $\xi_\kappa$ is given by

\begin{eqnarray*}
\xi_\kappa^{-2} = \frac{\alpha}{\pi \beta} \ln 2 
+ \frac{\kappa^2}{1 + \frac{\alpha \beta} {12 \pi}}
\end{eqnarray*}

\noindent
Integrating over the photon momentum $q$ at large distance $R$ 
the effective potential at finite temperature reads

\begin{eqnarray*}
V(R,\beta) &\simeq& - \frac{g^2}{2\pi} \int_0^\infty dq 
\frac{q J_0(qR)}{q^2 + \xi_\kappa^{-2}} 
\notag \\
&=& - \frac{\alpha}{N} \sqrt{\frac{\xi_\kappa}{8 \pi R}} 
e^{-R/\xi_\kappa}
\label{eqVR}
\end{eqnarray*}

\noindent
which shows that the stronger $\kappa$ the shorter the correlation length 
$\xi_\kappa$. Hence variations of $\kappa$ affects the correlation length 
between spinons. Moreover the variation of the flux through the square 
plaquette also affects the correlation length since the CS coefficient
is related to the flux through equation \eqref{eqKappaB}.
If the flux $\phi$ through the plaquette increases the 
correlation length $\xi_\kappa$ also increases, the larger $\kappa$ 
the shorter the interaction between spinons.

\subsection{Dynamical mass generation}

We show how the CS term affects the ``chiral'' restoring 
transition temperature of the dynamical mass generation.
The Schwinger-Dyson equation for the spinon propagator at finite temperature 
reads

\begin{eqnarray}
G^{-1}(k) &=& {G^{(0)}}^{-1}(k) 
\notag \\
&-& \frac{g}{\beta} \underset{\widetilde{\omega}_{F,n}}
{\sum} \int \frac{d^2 \vec{P}}{(2 \pi)^2} \gamma_\mu G(p) \Delta_{\mu \nu}
(k-p) \Gamma_\nu
\notag \\
\label{SchwingerDyson}
\end{eqnarray}

\noindent
where $p=(p_0=\widetilde{\omega}_{F,n},\vec{P})$,
$G$ is the spinon propagator, $\Gamma_\nu$ the spinon - photon
vertex which will be approximated here by its bare value $g \gamma_\nu$ and
$\Delta_{\mu \nu}$ is the dressed photon propagator \eqref{MCSpropagator}.
The second term in \eqref{SchwingerDyson} is the fermion self-energy $\Sigma$,
($G^{-1} = {G^{(0)}}^{-1} - \Sigma$).
Performing the trace over the $\gamma$ matrices in equation 
\eqref{SchwingerDyson} leads to a self-consistent equation for the self-energy

\begin{eqnarray}
\Sigma(k) = \frac{g^2}{\beta} \underset{\widetilde{\omega}_{F,n}}{\sum} 
\int \frac{d^2 \vec{P}}{(2 \pi)^2}
\Delta_{\mu \mu}(k-p) \frac{\Sigma(p)}{p^2 + \Sigma(p)^2}
\label{selfconsitent}
\end{eqnarray}

In the low energy and momentum limit $\Sigma(k) = m(\beta,\kappa) 
\simeq \Sigma(0)$. Equation \eqref{selfconsitent} simplifies to

\begin{eqnarray}
1 = \frac{g^2}{\beta} \underset{\widetilde{\omega}_{F,n}}{\sum} 
\int \frac{d^2 \vec{P}}{(2 \pi)^2}
\Delta_{\mu \mu}(-p) \frac{1}{p^2 + m(\beta,\kappa)^2}
\label{Mass1}
\end{eqnarray}

\noindent
If the main contribution comes from the longitudinal part 
$\Delta_{00}(0,-\vec{P})$ of the photon propagator \eqref{Mass1} goes 
over to

\begin{eqnarray}
1 &=& \frac{g^2}{\beta} \underset{\widetilde{\omega}_{F,n}}{\sum} 
\int \frac{d^2 \vec{P}}{(2 \pi)^2}
\notag \\
&&
\times
\left[
\left(\vec{P}^2 + \Pi_B(m=0) \right)
+ \frac{\left(\kappa \vec{P}\right)^2}{\left(\vec{P}^2 + \Pi_A(m=0)\right)} 
\right]^{-1}
\notag \\
&&
\times 
\frac{1}{\left(
\widetilde{\omega}_{F,n}^2 + \vec{P}^2 + m(\beta,\kappa)^2 \right)}
\end{eqnarray}

\noindent
where

\begin{eqnarray*}
\Pi_A \left(m=0 \right) &=& \Pi_1 \left(m=0\right) + \Pi_2 \left(m=0\right) \\
&=& \frac{\alpha P}{\pi} \int_0^1 dx \sqrt{x(1-x)} 
\tanh \beta P \sqrt{x(1-x)} \\
\Pi_B \left(m=0 \right) &=& \Pi_3\left(m=0 \right) \\
&=& 
\frac{\alpha}{\pi \beta} \int_0^1 dx \ln 2 \left(\cosh \beta P \sqrt{x(1-x)} 
\right)
\end{eqnarray*}

\noindent
Performing the summation over the modified fermion Matsubara frequencies 
$\widetilde{\omega}_{F,n} = \frac{2\pi}{\beta}(n+1/4)$ \cite{PRB,CondMat0602} 
the self-consistent equation takes the form

\begin{eqnarray}
1 &=& \frac{\left( \alpha/\Lambda \right)}{4 \pi N} \int_0^1 dP
\notag \\
&&
\times
P \tanh \left[\left( \beta \Lambda \right)
\sqrt{P^2 + \left(\frac{m(\beta,\kappa)}{\Lambda}\right)^2}
\right]
\notag \\
&&
\times
\left[ \left(P^2 + \frac{\Pi_B(m=0)}{\Lambda^2} \right)
+ \frac{\left(\frac{\kappa}{\Lambda} P\right)^2}
{\left(P^2 + \frac{\Pi_A(m=0)}{\Lambda^2} \right)} \right]^{-1}
\notag \\
&&
\times
\frac{1}{\sqrt{P^2 + \left(\frac{m(\beta,\kappa)}{\Lambda}\right)^2}}
\label{Mass2}
\end{eqnarray}

\noindent
As defined above $\alpha = g^2 N$ with $N=2$ since we have implemented the 
Popov-Fedotov procedure \cite{PRB,CondMat0602}. Here $\Lambda$ is the UV 
cutoff and can be identified as the inverse spin lattice spacing. Equation 
\eqref{Mass2} can be solved numerically.

\begin{figure}
\epsfig{file=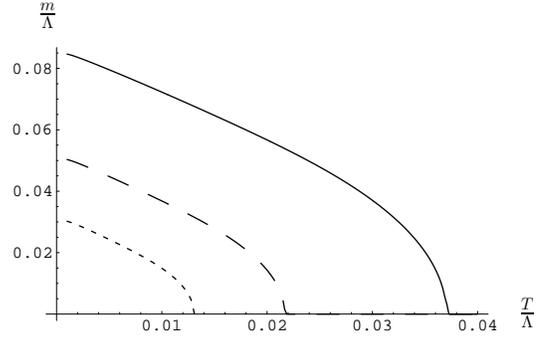,width=7cm}
\caption{Dependence of the dynamical mass $m(T)$ on the temperature. 
Full line : $\kappa/\Lambda = 0$. Long dashed line : $\kappa / \Lambda 
= 5.10^3$. Dashed line : $\kappa / \Lambda = 7.10^3$. All curve are obtained 
with $\alpha / \Lambda = 10^5$.}
\label{MassCSpf}
\end{figure}

\begin{figure}
\epsfig{file=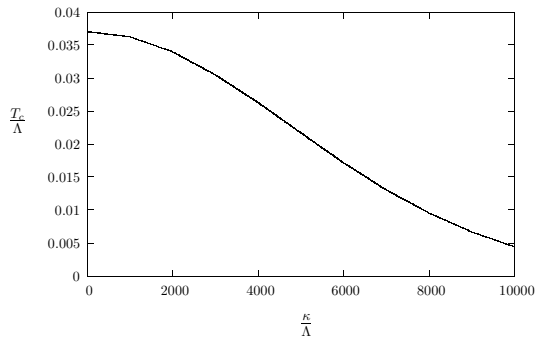,width=7cm}
\caption{Chiral symmetry transition temperature $\frac{T_c}{\Lambda}$
depending on the CS coefficient $\frac{\kappa}{\Lambda}$. Here
 $\frac{\alpha}{\Lambda} = 10^5$.}
\label{TcKappa}
\end{figure}

Figure \ref{MassCSpf} shows the dependence of the dynamical mass on the
temperature for different value of $\kappa / \Lambda$. 
This mass which is different from zero for low temperature $T$
vanishes at some temperature $T_c$ which depends on $\kappa$.
As $\kappa / \Lambda$ increases the chiral symmetry transition temperature
$T_c / \Lambda$ decreases following the relation
$\frac{T_c \left( \kappa \neq 0 \right) }{ T_c \left( \kappa = 0  \right) }
= e^{- a\left(\alpha / \Lambda \right) \frac{\kappa^2}{\Lambda^2} } $
where $a\left(\alpha / \Lambda \right)$ is a coefficient depending on 
$\alpha / \Lambda$. The behaviour for a fixed $\alpha$ is shown in 
Figure~{\ref{TcKappa}}.

One can understand the mechanism of chiral symmetry restoration as follows.
The photon (gauge field) gives a mass to the fermions (spinons) through a 
dynamical mass generation mechanism. When the temperature increases this 
mechanism is lowered by fluctuations, the fermions gain in mobility (the 
dynamical mass $m\left(\beta,\kappa\right)$ decreases).
This is similar to the situation in plasmas. When the temperature is high 
enough the charged particles composing the plasma
are considered as free particles. Below some temperature these
charged particles are screened thus their mass is renormalized and gets larger
than in the high temperature plasma.
The CS mass $\kappa$ contributes also to the photon mass, the interaction 
$V(R,\beta)$ between fermions is weakened as $\kappa$ increases for a fixed
length $R$, the correlation length $\xi_\kappa$ gets weaker and thus
the screening effect gets weaker.
Finally the chiral symmetry restoring temperature decreases with 
increasing $\kappa$ since the screening effect is smaller and thus fermions 
gain mobility, their mass term is renormalized to a smaller value. At zero 
temperature the dynamical mass decreases as $\kappa$ increases like 
$\frac{m(\kappa \neq 0)}{m (\kappa =0)}= e ^ {- \frac{4N}{N_c} \frac{\kappa^2}
{(\alpha/16)^2} }$ \cite{ItohKato}, thus proving that the screening effect is 
lower as the photon mass $\kappa$ in\-creases and reduces the dynamical mass 
of the spinon.

Going back to the Heisenberg model and looking for consequences of this
chiral symmetry restoration on the energy spectrum of spinons, one can see
that when the CS coefficient differs from zero
the gauge field $a_\mu$ gets a gap due to the mass term $\kappa$. 
As $\kappa$ goes to $\infty$ the spinon
energy gap $m(\beta, \kappa)$ decreases leaving a gapless 
spinon spectrum for a transition temperature $T_c$ going to zero.

For fixed $\alpha/\Lambda$ the knowledge of $T_c$ fixes $\kappa$ (Fig. 4) and 
consequently the generated mass $m(\beta,\kappa)$ (Fig. 3).

\section{Conclusions}

The low energy spectrum of the Heisenberg model describing a two dimensional
antiferromagnet quantum spin system with an $SU(2)$ symmetry has been mapped 
onto a $(2+1)$-dimensional quantum electrodynamics action 
with a $U(1)$ gauge field symmetry \cite{CondMat0602}.
In this framework we showed that the addition of a Chern-Simons term to the
Maxwell term in the pure gauge field theory of the $QED_3$ 
at finite temperature affects the spinon energy spectrum 
through a chiral symmetry restoration transition. The chiral symmetry
transition temperature $T_c$ above which the dynamical spinon mass is
equal to zero follows the relation 
$\frac{T_c \left( \kappa \neq 0 \right) }{ T_c \left( \kappa = 0  \right) }
= e ^{- a\left(\alpha / \Lambda \right) \frac{\kappa^2}{\Lambda^2} } $
where $a\left(\alpha / \Lambda \right)$ is a coefficient depending on 
$\alpha / \Lambda$ and $\Lambda$ the UV cutoff.

The effective potential between two spinons with opposite charge $g$
at finite temperature shows that the Chern-Simons term controls also the
screening of this interaction through the photon mass 
which is identified with the Chern-Simons coefficient $\kappa$. 
The correlation length $\xi_\kappa$ decreases showing
that the screening effect gets weaker as $\kappa$ increases.

The value of $\kappa$ can be controlled by fixing the flux through a plaquette 
going around neighbouring spin lattice sites. The gap in the spinon spectrum 
shrinks to zero with increasing $\kappa$ for a fixed temperature.
 
Hence the Maxwell-Chern-Simons term at finite temperature which is aimed to fix
the correct $U(1)$ gauge configuration provides an 
interesting way to control the chiral symmetry restoration temperature and the 
effective potential between spinons. The present study has been done in the 
framework of a non-compact theory. One may ask how it will be influenced by the 
presence of instantons in a compact description of the gauge field. This point 
related to the confinement/deconfinement problem is still under discussion
\cite{HandsKogutLucini,Herbut-02,NogueiraKleinert-05,Nazario}.

\begin{acknowledgments}
One of us (R.D.) would like to thank M. Rausch de Traubenberg for 
enlightening discussions and F. Stauffer for his encouragements.
\end{acknowledgments}

\section{Appendix \label{AppendixA}}

One may believe that a system at finite temperature breaks Lorentz invariance
since the frame described by the heat bath already selects out a specific 
Lorentz frame. However this is not true and one can formulate the statistical 
mechanics in a Lorentz covariant form \cite{Das}.

We consider a system in 2 space and 1 time dimension.
Define the proper $3$-velocity $u^{\mu}$ of the heat bath.
In the rest frame of the heat bath the three velocity
has the form $u^{\mu} = \left( 1,0,0 \right)$ and the inverse temperature
$\beta$ characterizes the thermal property of the heat bath.

Given the $3$-velocity vector $u^{\mu}$ one can decompose any three vector 
into parallel and orthogonal components with respect to the proper velocity
of the heat bath, the velocity $u^{\mu}$. In particular the parallel and 
transverse components of the three momentum $q^{\mu}$ with respect to 
$u^{\mu}$ read

\begin{eqnarray}
q^{\mu}_{\parallel} = \left( q . u \right) u^{\mu}
\end{eqnarray}
\begin{eqnarray}
\widetilde{q}^{\mu} = q^{\mu} - q^{\mu}_{\parallel}
\end{eqnarray}

\noindent
Similarly one can decompose any vector and tensor 
into components which is parallel
and transverse to a given momentum vector $q^{\mu}$

\begin{eqnarray}
\bar{u}_\mu &=& u_\mu - \frac{(q.u)}{q^2} q_\mu \\
\bar{\eta}_{\mu \nu} &=& \delta_{\mu \nu} - \frac{q_\mu q_\nu}{q^2}
\end{eqnarray}

\noindent
It is easy to define second rank symmetric tensors 
constructed at finite temperature from $q^{\mu}$, $u^{\mu}$ and 
$\delta_{\mu \nu}$ which are orthogonal to $q^{\mu}$

\begin{eqnarray}
A_{\mu \nu} &=& \delta_{\mu \nu} - u^{\mu} u^{\nu} 
- \frac{\widetilde{q}_\mu \widetilde{q}_\nu} {\widetilde{q}^2} \\
B_{\mu \nu} &=& \frac{q^2}{\widetilde{q}^2} \bar{u}_\mu \bar{u}_\nu
\end{eqnarray}

Since one considers a spin system
at finite temperature and ``relativistic'' covariance
should be preserved the polarization function may be put in the general 
form \cite{Das}

\begin{eqnarray}
\Pi_{\mu \nu} = \Pi_A A_{\mu \nu} + \Pi_B B_{\mu \nu}
\label{Chapter5eq23}
\end{eqnarray}

\noindent
and the Dyson equation \eqref{Dyson} can now be expressed in a covariant
form if one uses relation \eqref{Chapter5eq23}.

\end{document}